# An optics-free computational spectrometer using a broadband and tunable dynamic detector

**Authors**

Ling-Dong Kong[1], Qing-Yuan Zhao[1,2*], Hui Wang[1], Jia-Wei Guo[1], Hai-Yang-Bo Lu[1], Hao Hao[1], Shu-Ya Guo[1], Xue-Cou Tu[1,2], La-Bao Zhang[1,2], Xiao-Qing Jia[1,2], Lin Kang[1,2], Xing-Long Wu[3], Jian Chen[1,2*], and Pei-Heng Wu[1,2]

**Affiliations**

[1]Research Institute of Superconductor Electronics (RISE), School of Electronic Science and Engineering, Nanjing University, Nanjing, Jiangsu 210023, China.
[2]Purple Mountain Laboratories, Nanjing, Jiangsu 211111, China.
[3]National Laboratory of Solid State Microstructures and Department of Physics, Nanjing University, Nanjing, 210023, China

*Correspondence authors. Email: qyzhao@nju.edu.cn; chenj63@nju.edu.cn

**Abstract**

Optical spectrometers are the central instruments for exploring the interaction between light and matter. The current pursuit of the field is to design a spectrometer without the need for wavelength multiplexing optics to effectively reduce the complexity and physical size of the hardware. Based on computational spectroscopic results and combining a broadband-responsive dynamic detector, we successfully demonstrate an optics-free single-detector spectrometer that maps the tunable quantum efficiency of a superconducting nanowire into an ill-conditioned matrix to build a solvable inverse mathematical equation. Such a spectrometer can realize a broadband spectral responsivity ranging from 660 to 1900 nm. The spectral resolution at the telecom is 6 nm, exceeding the energy resolving capacity of existing infrared single-photon detectors. Meanwhile, benefiting from the optics-free setup, precise time-of-flight measurements can be simultaneously achieved. We have demonstrated a spectral LiDAR with 8 spectral channels. This work provides a concise method for building multifunctional spectrometers and paves the way for applying superconducting nanowire detectors in spectroscopy.

# MAIN TEXT

## Introduction

Optical spectroscopy is a widely used technique to investigate wavelength-dependent interactions between light and matter across countless cutting-edge research fields(*1, 2*) and industrial processes(*3*), where spectrometers are the central instruments. In conventional spectrometers, the incident light passes through spectral multiplexing optics to separate spectral components either at different times(*4*) or different locations(*5*). Therefore, the information about the spectral intensity distribution can be known after measuring each spectral component by a photodetector or a detector array. Although the measurement step is straightforward, distinguishably separating wavelengths over spatial or temporal dimensions requires high-quality grating, prisms, or tunable bandpass filters. Thus, the bandwidth and resolution of a conventional spectrometer are limited by these complex wavelength multiplexing optics, especially for a system requiring a simplified configuration.

Computational spectrometers use a different approach to get the spectral information of incident light and give much more freedom in designing optics and choosing detectors, resulting in a miniaturized footprint(*5*). Different from a conventional spectrometer, the spectral component is no longer required to be derived from a single-shot measurement. Instead, each measurement could be a superposed response from a broadband incident light with an unknown spectrum $x$. By conducting a group measurement $y$, $x$ can be reconstructed by computing the inverse problem of $x = \Phi y$, where $\Phi$ is a calibrated two-dimensional matrix containing the spectral information of the system. In a computational spectrometer, the key is to build a linearly independent $\Phi$ and optimize the reconstruction algorithm accordingly.

Previous works for building $\Phi$ are based on a filter array of varied optical transmission(*6-9*), a tunable optics by time multiplexing(*10, 11*), or a detector array of varied spectral responsivity(*12-14*) (see Fig. S9 for a comparison between different approaches). However, to our knowledge, it has not been demonstrated to realize a computational spectrometer by using a single detector solely. Although with a photon-counting detector of energy resolving capacity, a spectrometer can be simplified to a single detector and wavelength multiplexing optics can be removed, the spectral resolution is limited by the sensitivity of the analog readout from the detector. For instance, superconducting transition edge sensors (TESs) that are used widely in X-ray spectroscopy have single-photon sensitivity and the output amplitude is proportional to the energy of photon energy(*15*). Thus, TESs can be used as a single-detector spectrometer by counting the photon counts and measuring the pulse amplitude. However, the energy resolution of a TES is limited to be ~0.11 eV(equivalent to a 187 nm wavelength resolution between 1363 nm and 1550 nm) for infrared photons(*16*). Time-of-flight spectrometers use a time-multiplexing way to separate light so that one detector of low timing jitter can read out the wavelength of a single photon (*4*). However, this method can only work for synchronized pulsed light and need strong dispersive media, e.g. a km-long fiber.

Here, we build an optics-free, single-detector spectrometer (SDS) based on a dynamic detector and a computational spectroscopy method as shown in Fig. 1A. The spectral responsivity of this dynamic detector depends on both the wavelength and the bias, which naturally contributes to a programmable $\Phi$. In this work, we choose a superconducting nanowire single-photon detector (SNSPD) as an example. SNSPD is a photon-counting detector. They can have spectral responsivity from ultraviolet(*17*) to mid-infrared(*18, 19*), single-photon sensitivity, high dynamic range, and can be integrated on-chip with nanophotonic circuits(*20*). It is demonstrated both experimentally and theoretically that the quantum efficiency of an SNSPD has a strong dependence on its bias current $I_B$ and the wavelength $\lambda$ of the incident light(*18, 19*). This nonlinear dependence was noticed and researchers demonstrated to use it for spectroscopy a few years after SNSPD was developed(*21*). Although their results were preliminary, the idea that SNSPD was capable to build spectrometers in an unconventional way was striking. By sweeping the SNSPD's bias current and collecting the photon counts, the forward mathematical equations can be built. Accompanied with a computational algorithm to solve the inverse problem, the unknown spectrum is then reconstructed. Since no wavelength multiplexing optics are required, the proposed dynamic SDS fully utilizes the broadband response of the detector.

Experimentally, we have demonstrated both monochromic and broadband spectral spectra from 660 nm to 1900 nm. A potential extension of the range to X-ray(*22*), ultraviolet(*17*), and longer wavelength(*19*) can be expected reasonably based on the recent developments of superconducting nanowire detectors. As electrical modulation is used, a large-size $\Phi$ can be generated programmatically without the need for a complex readout for detector arrays proposed in previous works(*6-9, 12-14*). Consequently, a high spectral resolution of sub-10 nm is obtained at the telecom band. As the configuration of the spectrometer no longer depends on wavelength multiplexing optics or complex detector arrays, the SDS can be combined inherently with existing systems. For instance, a LiDAR or a fluorescence lifetime measurement system can gain additional spectral information straightforwardly. To demonstrate such an advantage, we build a spectral LiDAR, which can provide intensity image, depth image, and spectral images of eight channels all in one system.

**Results**

We fabricated a polarization-insensitive SNSPD (Fig. 1B). The system spectral responsivity including the tunable quantum efficiency of the nanowire and the coupling efficiency at the $i^{th}$ bias current $I_B$ was calibrated to build the responsivity matrix $\Phi$, which is shown in Fig. 1C. $\Phi$ is a $n \times n$ matrix, where $n$ is the number of spectral bands. Each row of the matrix $\Phi$ represents a system spectral responsivity at different bias currents. Figure 1D shows different spectral responsivities at three selected currents, from which we can see that dynamic tuning of spectral responsivity of the superconducting nanowire is obvious over a wide wavelength range simply by changing $I_B$. Under low bias current, photons of short wavelength can be detected with higher probability. When the bias current increasing, the responsive spectrum extends to a longer wavelength. For each wavelength, i.e., in each column of the matrix $\Phi$, the photon counting rate (PCR) versus bias current is a sigmoidal curve as shown in Fig. 1E. For shorter wavelength light, the sigmodal curve goes to saturate at an earlier inflection point of the bias current and the growth is faster than the light of longer wavelength. During the measurement of the unknown spectrum, the measured PCR is the linear superposition of a series of PCR curves at the corresponding wavelengths with different weights. The dependence of the dark counts on bias current is also calibrated, which is shown in Fig. 1F. The dark count rate of the SNSPD is below 1000 cps, which is an ultralow value compared to the PCR (~$10^5$ cps).

Due to the in-situ tunable spectral responsivity of SNSPDs, the SDS can be applied to different spectral ranges and resolutions for various applications using the same detector. In this paper, we calibrated three different responsivity matrices, which were $\boldsymbol{\Phi_{brd}}$ (Fig. 1C) for broadband measurement, $\boldsymbol{\Phi_{tel}}$ (Fig. 3A) for resolution measurement at telecom band, and $\boldsymbol{\Phi_{img}}$ (Fig. 4F) for spectral imaging experiments.

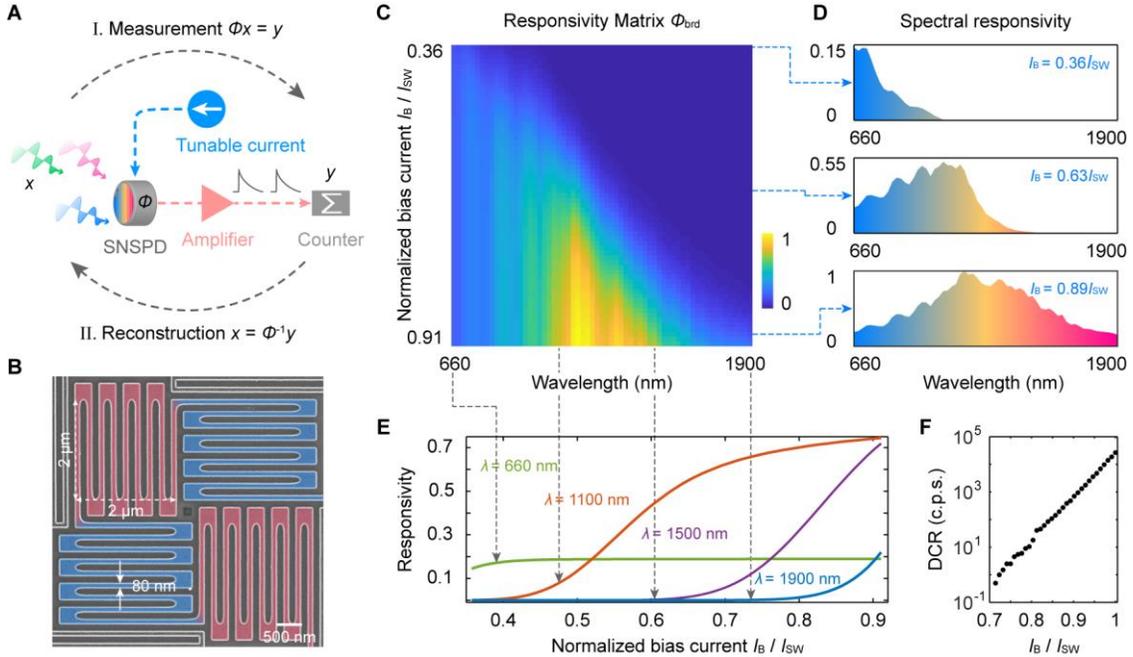

**Fig. 1. Superconducting nanowire single-detector spectrometer.** (**A**) Operating principle diagram. (**B**) A scanning electron micrograph of the SNSPD, which was meandered in perpendicular directions to reduce photon polarization sensitivity (painted with two different colors). (**C**) The broadband calibrated responsivity matrix $\boldsymbol{\Phi_{brd}}$ as a function of wavelength and bias current. The data is normalized to the maximum responsivity. (**D**) The dependence of $\boldsymbol{\Phi_{brd}}$ on $I_B$ at fixed $\lambda$. (**E**) The dependence of $\boldsymbol{\Phi_{brd}}$ on $\lambda$ at fixed $I_B$. (**F**) Dark count rate versus bias current.

With the calibrated spectral responsivity matrix $\boldsymbol{\Phi}$, the SDS measured spectral information through two steps—measurement and reconstruction, as illustrated in Fig. 1A. First, the photon count rates (PCR) $c_i$ were collected at the $i^{\text{th}}$ bias current $I_B^i$. By sweeping the current incrementally with $n$ steps, the direct measurement vector $\boldsymbol{y}$ is obtained. Second, the reconstruction algorithm aims to reliably estimate the target spectrum $\boldsymbol{x}$ from $\boldsymbol{y}$ and $\boldsymbol{\Phi}$. Notably, the condition number of the matrix $\boldsymbol{\Phi}$ is extremely large ($1\times10^7$ for $\boldsymbol{\Phi_{brd}}$, $5\times10^7$ for $\Phi_{\text{tel}}$ and $4\times10^4$ for $\Phi_{\text{img}}$), which implies the spectral reconstruction of $x$ is an ill-posed problem(*5*). For ill-posed problems, the results are sensitive to perturbations of data due to errors. Thus, the regularization algorithm is included. Here, we used a truncated generalized singular value decomposition (TGSVD) method to reduce the influence of perturbations during the reconstruction(*23*), which is a popular solver for the regularized solution of ill-posed inverse problems in imaging processing and machine learning. Moreover, we minimized the measurement uncertainty by stabilizing the operating temperature with a fluctuation of less than $\pm0.15$ mK, amplifying the detector's output with a cryogenic amplifier to reduce electrical readout noise, and accumulating photon counts to reduce photon counting noise.

Following the above workflow, we demonstrated the capability and performance of the computational single-detector spectrometer by measuring and reconstructing different incident-light spectra from both monochromatic light sources and broadband light sources. A multimode fiber (MMF, core size 62.5 μm) was used to couple light to the SNSPD. The MMF can degenerate the coherence and scramble the polarization of the incidence light and its spectral characteristics were included in the calibration matrix. Therefore, our experiments worked for different light sources (supercontinuum light source, super-luminescent diode, and fiber-based mode lock laser). However, as the active area of the detector was smaller compared to the speckle pattern output from the MMF(*24*), the nonuniformity of light intensity of the speckle pattern resulted in irregular peaks and oscillations in the calibration matrix. Such fine features demand that the system should be calibrated precisely and maintained stably during measurements. More discussions about these fine features and effects of the coherence of light are given in the supplementary information.

The broadband performance of the SDS was shown in Fig. 2. We used a broadband supercontinuum source (YSL SC-PRO-M) with an acoustic-optical tunable filter (AOTF, bandwidths are between 4 nm~16 nm) to generate swept light. A power meter and counter are used to measure the incident light power and the PCRs, from which a normalized (photon rates per watt) $\Phi_{\text{brd}}$ can be built. $\Phi_{\text{brd}}$ has a size of $63 \times 63$ with bias current $I_B$ ranging from 4 μA to 10.2 μA and wavelength $\lambda$ ranging from 660 nm to 1900 nm. We used the same broadband source and filters in calibration to generate narrowband light from visible to infrared and measured their photon counts versus current (Fig. 2A). The reconstructed spectra are shown in Fig. 2B. The center frequencies agreed well with the referenced centers set by the AOTF. For testing the broadband reconstruction performance. The output of the same broadband light source used in calibration passed through two different optical filters to generate two different broadband signals. Fig. 2C shows the measured PCRs. Figure 2D and 2E show the corresponding reconstructed spectra, which agree with the reference data measured by a commercial spectrometer. These results indicate that the SDS works for a broad range of both narrowband and broadband signals.

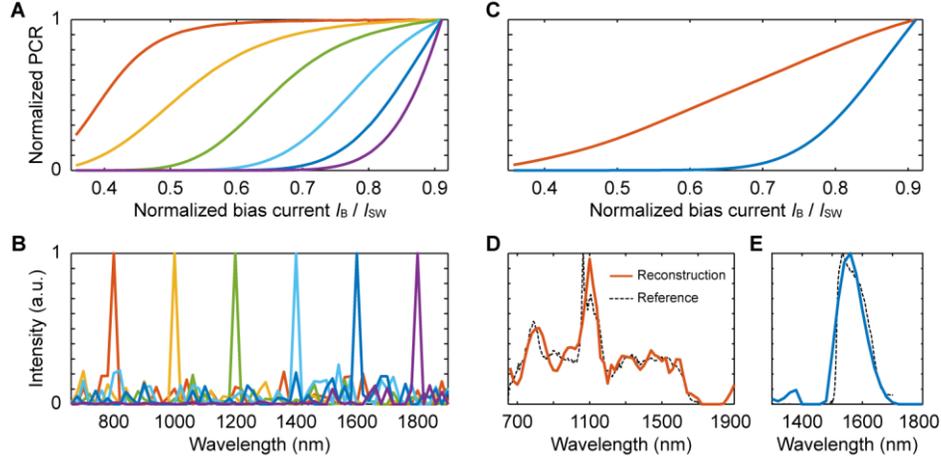

**Fig. 2. Results of broadband spectrum reconstruction.** (**A**) Measured PCRs for a series of narrow-band light from 800 nm to 1800 nm. (**B**) Reconstructed spectra from the PCRs (marked as the same color shown in (A)). (**C**) Measured PCR curves for the output from a supercontinuum source through a 650 nm long-pass filter (orange) and a 1750 nm **band**-pass filter with a bandwidth of 500 nm (blue). (**D, E**) Reconstructed spectra from the PCRs in (C). The reference spectral measured by a commercial spectrometer is shown as the black curve.

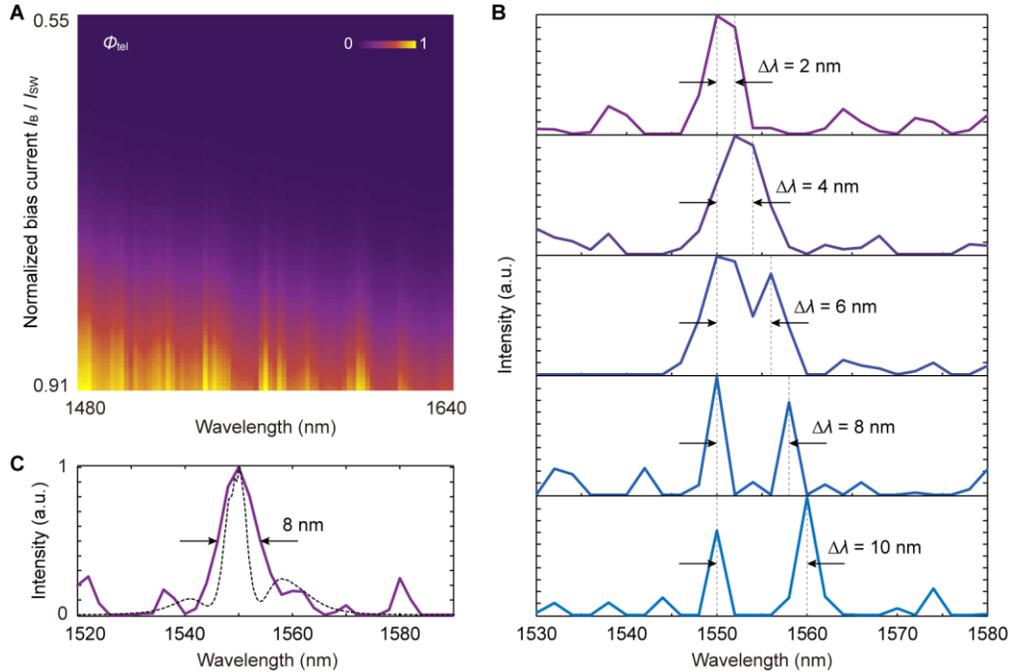

**Fig. 3. Spectral resolution measured at the telecom band.** (**A**) A refined responsivity matrix $\Phi_{tel}$ calibrated at the telecom band. (**B**) Reconstructed spectra of two monochromatic light sources separated with a wavelength distance from 2 nm to 10 nm. The dotted lines mark the locations of the dual incident wavelengths. (**C**) The reconstructed spectrum of a femtosecond laser. Reference spectra are shown as the black curves.

The spectral resolution of the single-detector spectrometer was measured at the telecom band. A finer spectral responsivity matrix $\boldsymbol{\Phi}_{\text{tel}}$ (Fig. 3A) was calibrated with a tunable laser (Santec TSL 710). $\boldsymbol{\Phi}_{\text{tel}}$ had a size of 81 × 81 ($I_B$ from 0.55$I_{SW}$ to 0.91$I_{SW}$ and $\lambda$ from 1480 nm to 1640 nm). As shown in Fig. 3B, when two incident monochromatic lights were separated by 6~8 nm, the spectrum reconstruction can still be done and spectral peaks were resolved. We also measured the spectrum of a mode-lock femtosecond laser (Calmer FPL02CFFPM). As shown in Fig. 3C, the laser has a spectrum of 5 nm bandwidth defined at the full-width-at-half-magnitude (FWHM), while the SDS gives a reconstructed spectrum of 8 nm bandwidth. Considering the measured spectrum is a convolution of the instrument response function (IRF) of the SDS and the spectrum of the signal, the resolution defined by the FWHM of the IRF is 6.2 nm, which agrees with previous measurements of two separate signals. Further improvement of the resolution is possible by optimizing the calibration precision, measurement accuracy, and reconstruction algorithm.

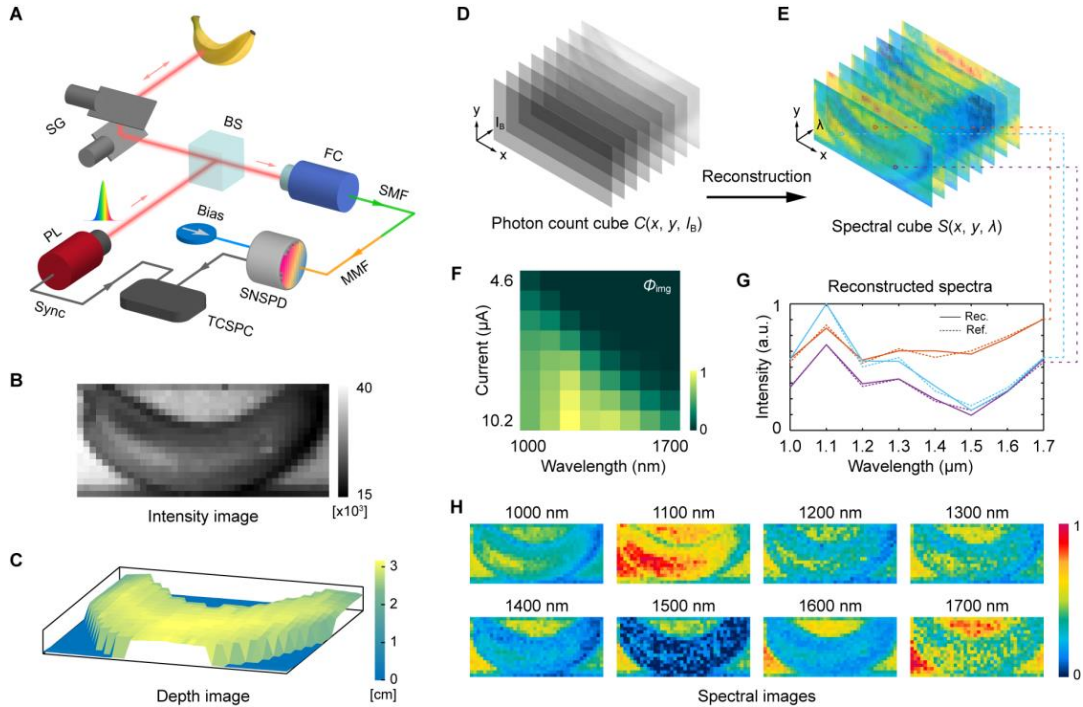

**Fig. 4. Spectral LiDAR by adopting the SDS method.** (**A**) The raster scanning LiDAR setup. PL: pulsed laser, FC: fiber coupler, BS: beam splitter, SG: scanning galvo. (**B**) Intensity image using the photon count data. (**C**) The depth image based on the ToF data. (**D**) Data cube of the photon count at scanned locations and bias currents. (**E**) Data cube of the spectral images after reconstruction. (F) The calibrated responsivity matrix $\boldsymbol{\Phi}_{\text{img}}$. (**G**) Reconstructed spectra at three positions selected from the scene. The orange line represents one position at the plastic plate (background). The blue and purple lines represent two locations at the object, which is a ripe banana. The reference spectrum was measured by sending the narrow-band light individually. (**H**) Spectral images of the scene at different wavelengths.

Benefiting from the minimum configuration of the SDS, existing systems that using SNSPD as the detector can obtain additional spectral sensitivity straightforwardly. Because of the low timing jitter and high efficiency, SNSPDs have been applied in long-range LiDARs(*25, 26*). Here, we built a single-photon LiDAR and upgraded it to a spectral LiDAR by introducing the SDS method. As shown in Fig. 4, the system used the typical coaxial raster scanning configuration. The illumination source contained 8 narrow-band light ranging from 1000 nm to 1700 nm and was modulated into a 280 ps wide pulse with a repetition rate of 10 MHz to mimic a broadband light source. At each scanning position (*x*, *y*), the detector was bias at 8 different bias current $I_B$ ranging from 4.6 µA to 10.2 µA. At the $i^{th}$ bias, the time-correlated counting histogram was recorded by a TCSPC, from which we can get the total counts $C_i(x, y, I_B^i)$ and the time-of-flight (TOF) value $T_i(x, y, I_B^i)$. At the maximum bias current where the detector had the lowest timing jitter of 18 ps (Fig. S2B), we generated the intensity image based on the reflected photon flux (Fig. 4B) and the depth image based on the TOF value (Fig. 4C). By adopting the SDS method with the calibrated responsivity matrix $\boldsymbol{\Phi}_{\mathbf{img}}$ of an 8×8 size (Fig. 4F) and the data cube of $C(x, y, I_B)$ (Fig. 4D), we can have the multispectral data cube $S(x, y, \lambda)$ (Fig. 4E). As shown in Fig. 4G, the reflectivity spectra were reconstructed successfully and agreed with the reference at positions from the object and background. Spectral images at different wavelengths were shown in Fig. 4H, from which we can have the spectral characteristics of the object (a ripe banana) and distinguish it clearly from the background.

**Discussion**

As a reconstruction method is used, uncertainties in calibration and measurement will cause errors. And this error will be amplified in the reconstrued spectrum, deteriorating the performance. This is a common problem for solving ill-posed inverse problems and would limit the application of SDS, which can be overcome by a joint development of the device design and algorithm. Speed is another concern in the SDS, which can be improved by using a faster detector or adopting a compressive algorithm. If the detector was optimized at a higher count rate, i.e. ~100 MHz based on the reported maximum count rate of an SNSPD, the acquisition time could be estimated to speed up to several seconds for having the same spectrum reconstruction quality. Inspired by the compressive sensing method, for sparse spectra, we can use fewer sampling currents to achieve the reconstruction with slightly reduced accuracy. We developed a compressive spectral reconstruction algorithm by exploiting sparse representation based on various Gaussian basis(*12, 13*) and the smoothing feature of natural signals (supplementary ST3). For the sparse spectrum shown in Fig. 2E, when the sampling rate decreased to 10%, the reconstructed spectrum agreed with the referenced spectrum as shown in Fig. S8. The peak signal-to-noise ratio (PSNR) of the reconstructed spectrum was reduced from 24.72 dB to 20.46 dB. If the spectra were sparsely represented in a more sophisticated dictionary, an optimized reconstruction of an even higher-compression ratio can be expected.

In summary, we have demonstrated a single-detector spectrometer that using only one photodetector in the absence of any wavelength multiplexing optics. Due to the avoidance of optics and detector arrays, the single-detector spectrometer can reduce the complexity of configuration to the minimum. Many existing systems that using superconducting nanowire single-photon detectors (SNSPDs) or other dynamic detectors can gain spectroscopy function by integrating the SDS method without changing much of the hardware. The SDS takes full advantage of the detector so that high spectral resolution and wide spectral range can be achieved together compared to other computational spectrometers (supplementary Table S1).

From the perspective of developing advanced detectors, our results indicate that although a single-photon detector, such as an SNSPD, has no energy resolving capacity from a single-shot measurement, by programming the measurements and combining a computational algorithm, the spectrum of the incident light can be measured computationally. This is extremely important to various dynamic detectors made from semiconductors(*27, 28*), superconductors(*18, 19*), and 2D materials(*29*) that have already in-situ tunable responsivity. By adding parametric sweeping measurements and post signal processing, the existing hardware can gain additional spectrum information, enabling a multifunctional system that would be used in astronomical observation, biological analysis, and remote sensing.

## Materials and Methods

### Device fabrication

6.6-nm-thick NbN film was deposited on a 4-inch silicon wafer with a 268-nm-thick thermal oxide layer using reactive sputtering. The NbN film had a critical temperature $T_c$ = 7.6 K. We spun a 76-nm-thick positive-tone resist (950PMMA A2) and then baked it at 150 °C on a hot plate for 3 min. The device structure was patterned using 100 kV electron-beam lithography. The pattern was transferred into the NbN layer using $CF_4$/Ar reactive-ion etching at 50 W for 70 s. Finally, we spun a new layer of 90-nm-thick PMMA as a protective coating.

### Experimental setup of the SDS

All measurements were carried out in a pulse-tube cryocooler with a PID control to stabilize the temperature at 1.654 K (Fig. S1). Output pulses of the SNSPD were amplified by a cryogenic amplifier (Cosmic Microwave, CITLF3) with an integrated Bias-T. The amplifier was mounted on the 1.6 K stage of the cryocooler and biased with a supply voltage of 1.6 V for low dissipative power (about 5.6 mW). The cryogenic amplifier provided a high signal-to-noise ratio (SNR) of electrical pulses (Fig. S2), which generated a valid registration of each output pulse by a universal counter even when the SNSPD was biased at a relatively low current.

The output light from the tunable laser was partially polarized, which would affect the measurement stability of the photon count rate. To reduce the influence of polarization, on one hand, we designed a polarization-insensitive SNSPD with an orthogonal structure. On the other hand, a 4-m-long silica multimode fiber (MMF) was exploited to couple the incident light to the detector, which randomized the polarization of light (a previous work demonstrated that the ellipticity of light decreased by 70 times with a 4-m-long silica multimode fiber(*18*)).

During the calibration process, the SNSPD chip was illuminated from the top with narrow-band light of different wavelengths, whose power was measured by a power meter. The photon count rate (PCR) was measured by a counter with a counting time of 10 seconds. The power at each wavelength was adjusted to make the maximum PCR less than $1\times10^6$ count/second so that the device operated in the linear counting regime.

**Measurement model of the SDS**

The measurement can be described by

$$c_i = \int_{\lambda_{\min}}^{\lambda_{\max}} s(\lambda)\varphi_i(\lambda)d\lambda \tag{1}$$

where $c_i$ is the, $s(\lambda)$ is the spectrum of the incident light that is unknown with a wavelength range from $\lambda_{\min}$ to $\lambda_{\max}$, $\varphi_i(\lambda)$ is the spectral responsivity of the detector at different bias currents. By discretizing $s(\lambda)$ and $\varphi_i(\lambda)$ into $s(k)$ and $\varphi_i(k)$ ($k$ = 1, 2, 3, …, n), we obtain

$$c_i = \sum_{k=1}^{n} \varphi_i(k)s(k) \tag{2}$$

Assuming

$$\boldsymbol{\Phi} = \begin{bmatrix} \varphi_1(1) & \varphi_1(2) & \cdots & \varphi_1(n) \\ \varphi_2(1) & \varphi_2(2) & \cdots & \varphi_2(n) \\ \vdots & \vdots & \ddots & \vdots \\ \varphi_m(1) & \varphi_m(2) & \cdots & \varphi_m(n) \end{bmatrix}, \boldsymbol{x} = [\,s(1), s(2), \cdots, s(n)\,]^{\mathrm{T}}, \text{and } \boldsymbol{y} = [c_1, c_2, \cdots, c_m]^{\mathrm{T}},$$

the measurement process can be modeled discretely as

$$\boldsymbol{\Phi x} = \boldsymbol{y} \tag{3}$$

The directly measured responsive matrix $\boldsymbol{\Phi}$ is a production of multiple matrices:

$$\boldsymbol{\Phi} = \boldsymbol{QAPT} \tag{4}$$

where $\boldsymbol{T}$, $\boldsymbol{P}$, and $\boldsymbol{A}$ are diagonal matrices with a size of $n \times n$ ($n$ is the number of spectral bands). $\boldsymbol{T}$ is the spectral transmission of a series of optics, $\boldsymbol{P}$ transfers light intensity to photon number, $\boldsymbol{A}$ is the spectral absorption of the device, and $\boldsymbol{Q}$ is the intrinsic quantum efficiency of the device. Among these four parameters, only $\boldsymbol{Q}$ depends on the bias current while $\boldsymbol{A}$, $\boldsymbol{P}$ and $\boldsymbol{T}$ are current independent.

**Supplementary Materials**
Supplementary Text
Figs. S1 to S9
Table S1
References (*1-14*)

**Acknowledgments**

We thank the other members of RISE for assistance in nanofabrication, measurements, and providing instruments. **Funding:** This work was supported by the National Key R&D Program of China Grant (2017YFA0304002), the National Natural Science Foundation (Nos. 61521001, 62071214, 61801206, 61571217, 61801209 and 11227904), the Recruitment Program for Young Professionals, the Fundamental Research Funds for the Central Universities, the Priority Academic Program Development of Jiangsu Higher Education Institutions (PAPD), the Qing Lan Project and the Jiangsu Provincial Key Laboratory of Advanced Manipulating Technique of Electromagnetic Waves. **Author contributions:** L.-D.K. and Q.-Y.Z. conceived the initial idea. L.-D.K. fabricated the device, performed the experiments, analyzed data, and developed the reconstruction program. H.W., H.L., H.H., and J.C. helped with cryogenic setup. J.-W.G simulated the absorbance of the SNSPD. S.-Y.G., X.-C.T., L.-B.Z., X.-Q.J., and L.K. helped with device fabrication. Q.-Y.Z., C.J., and P.W. supervised the work. L.-D.K., Q.-Y.Z., C.J., and X.-L. W. discussed the results and wrote the manuscript. **Competing interests:** The authors declare no competing financial interests. **Data and materials availability:** All data needed to evaluate the conclusions in the paper are present in the paper and the Supplementary Materials. Additional data are available from the corresponding author upon reasonable request.